\documentclass[useAMS,usenatbib]{mn2e}

\usepackage{threeparttable}
\usepackage{pdflscape}
\usepackage{graphicx}
\usepackage{longtable}
\usepackage{url}

\title[H$_{2}$O maser astrometry: S255IR-SMA1]{H$_{2}$O masers in a jet-driven bowshock: Episodic ejection from a massive young stellar object}
\author[R. A. Burns]{R. A. Burns$^{1}$\thanks{E-mail:
RossBurns88@googlemail.com}, T. Handa$^{1}$, T. Nagayama $^{2}$, K. Sunada$^{2}$, and T. Omodaka$^{1}$\\
$^{1}$Graduate School of Science and Engineering, Kagoshima University, 1-21-35 K\^orimoto, Kagoshima 890-0065, Japan\\
$^{2}$Mizusawa VLBI Observatory, National Astronomical Observatory of Japan,
2-12 Hoshigaoka-cho, Mizusawa, Iwate 023-0861, Japan}
\begin{document}

\date{Accepted 1988 December 15. Received 1988 December 14; in original form 1988 October 11}

\pagerange{\pageref{firstpage}--\pageref{lastpage}} \pubyear{2002}

\maketitle

\label{firstpage}

\begin{abstract}
We report the results of VERA multi-epoch VLBI 22 GHz water maser observations of S255IR-SMA1, a massive young stellar object located in the S255 star forming region.
By annual parallax the source distance was measured as D $= 1.78^{+0.12}_{-0.11}$ kpc and the source systemic motion was ($\mu_{\alpha}\cos\delta$, $\mu_{\delta}$) = ($ -0.13\pm0.20$, $-0.06\pm0.27$) mas yr$^{-1}$. 
Masers appear to trace a U-shaped bow shock whose morphology and proper motions are
well reproduced by a jet-driven outflow model with a jet radius of about 6 AU.
The maser data, in the context of other works in the literature, reveal ejections from S255IR-SMA1 to be episodic, operating on timescales of $\sim 1000$ years.

\end{abstract}


\section{Introduction}

The rarity, deeply embedded nature and large distances to massive young stellar objects (MYSOs) renders them difficult to observe and as a result their formation process, and especially the distinction from low-mass star formation, has yet to achieve consensus among astronomers.
As the number of observational studies increases we begin to find MYSOs which exhibit many of the features commonly associated with low-mass star formation, such as circumstellar disks \citep{Beltran04,Hirota14}, collimated outflows \citep{Beu02,Davis04} and jet rotation \citep{Burns15a}. Distilling the distinction between low and high mass star formation is therefore of high priority in the search for a theory of massive star formation.

\citet{Corc98} show that, in low-mass YSOs, the L$_{\rm IR}$ excess, a tracer of accretion activity, scales linearly with outflow luminosity - a relationship indicative of the typical disk-jet-outflow process ubiquitous to low-mass star formation. This relationship was found to extend unbroken from L$_{\rm bol} = 10^{-1}$ to $10^5$ L$_{\odot}$ \citep{Garatti15}, suggesting that stars in this luminosity range may be formed by a common process; that low mass star formation mechanisms can be `scaled up' to a few tens of solar masses.

In the case of low-mass stars, outflows are thought to be predominantly driven by disk-launched collimated jets, with some contribution form a disk wind \citep{Arce07}. In this scenario linear momentum from the high velocity jet seeps into the ambient gas, entraining it and producing an extended, low-velocity molecular outflow.

Whether or not the outflows of low- and high-mass stars are produced in the same way remains to be firmly established - similarities are seen \citep{Zhang02} however the often cited lower degree of collimation \citep{Wu04} in outflows from massive stars may indicate a different origin, such as a radiation driven outflow brought about by the flashlight effect in wind-blown cavities \citep{Zinnecker07}.
Uncovering the driving mechanism of outflows in MYSOs is therefore an accessible point of comparison between high-mass and low-mass star formation.



Episodic accretion is also emerging as a vital component of star formation, both for low mass \citep{Zu09,Stamatellos11} and very high mass primordial stars \citep{Hosokawa15}. The accretion history of a YSO can be inferred from its ejection history - as such, investigating outflows also exposes the accretion behaviour of MYSOs.

S255IR-SMA1 is the brightest source of molecular line emission in the S255IR star forming region (\emph{see} \citealt{Zin15}), which is sandwiched between two evolved H$\mathrm{II}$ regions, S255 and S257. There also exists a compact source of continuum emission in both millimeter \citep{Wang11,Zin15} and centimeter \citep{Zin15,Reng96}. \citet{Zin15} estimate the mass of the embedded star as M $\simeq 20$ M$_{\odot}$ based on a spectral energy distribution compiled from \citet{Oj11} and \citet{Zinchenko09}.

VLBI maser observations of S255IR-SMA1 were carried out by \citet{Goddi07} in 2005 using the Very Long Baseline Array (VLBA) which revealed the presence of clusters of 22 GHz H$_{2}$O masers, at close proximity ($<100$ AU) to the MYSO - suggesting the possible presence of a new jet ejection event aligned geometrically with bipolar molecular outflows reported by \citet{Wang11} and \citet{Zin15}. As such, S255IR-SMA1 is an ideal target to study the relationship between outflows and jets - in addition to episodic behaviour - in an MYSO.

Five years after the observations of \citet{Goddi07} we pursue new VLBI observations aimed at measuring the annual parallax, distribution and proper motions of water masers in S255IR-SMA1. We use our data to investigate the nature and driving mechanism of the ejections from S255IR-SMA1.











\section{Observations and Data Reduction}
\label{obs}

All observations were carried out using VLBI Exploration of Radio Astrometry (VERA).
Observations of S255IR-SMA1, and the continuum positional reference source J0613+1708 (separation $= 0.87^{\circ}$, PA $= 169^{\circ}$) were made simultaneously by utilising the dual-beam capabilities of VERA \citep{dual}. This removes the need to slew antennae and interpolate atmospheric phase solutions between sources - as is required for \emph{fast-switching} VLBI phase referencing. As such, dual-beam observations achieve excellent suppression of the dynamic troposphere phase contribution, which is known to be the dominant source of astrometric phase error at 22 GHz \citep{Asaki07}.


Typical observing sessions were $\sim$8 hrs long, providing $\sim$2.5 hrs on-source time and good \emph{uv}-coverage. The synthesised beam was typically $1.4\times0.9$ mas, PA $=-49^{\circ}$.
Intermittent observations of BL Lac, DA55 or 3C84 were made every 1.5 hrs for bandpass and group delay calibration.
The continuum source J0613+1708 is found in the VLBA calibrator list \citep{VLBA2} and had an unresolved $K$-band flux of $\sim$40 mJy in our observations with VERA.

Left-handed circular polarisation signals were recorded to magnetic tapes at each VERA station, sampled at 2-bit quantisation, and filtered with the VERA digital filter unit \citep{Iguchi05}.
Signal correlation was carried out using the Mitaka FX correlator \citep{Chikada}.
Individual source phase tracking centers were set to
$(\alpha, \delta)_{\mathrm{J}2000.0}=(06^{\mathrm{h}}12^{\mathrm{m}}54^{\mathrm{s}}.00640929$,
+17$^{\circ}$59'22".95890) and
$(\alpha, \delta)_{\mathrm{J}2000.0}=(06^{\mathrm{h}}13^{\mathrm{m}}36^{\mathrm{s}}.360073$,
+17$^{\circ}$08'24".94542) for S255IR-SMA1 and J0613+1708, respectively.
The total post-correlation bandwidth of 240 MHz was divided into 16 intermediate frequency (IF) channels, with 15 IFs being allocated to continuum sources and 1 IF allocated to the maser emission.
The maser IF, assuming a rest frequency of 22.235080 GHz, was correlated in `zoom band' mode giving 8 MHz bandwidth and a channel spacing of 15.63 kHz, corresponding to a velocity resolution of 0.21 km s$^{-1}$. Fourteen of the 15 IFs allocated to continuum sources had bandwidths of 16 MHz and channel spacings of 125 kHz. The 15th IF allocated to continuum sources was correlated at the same bandwidth and resolution as the maser zoom band. In order to make uniformity in the continuum source IFs the 16 MHz bandwidth IFs were halved and combined with the zoomed 8 MHz bandwidth IF, allowing all 15 IFs to be merged.

\begin{table}
\scriptsize
\caption{Summary of observations\label{obs}}
\begin{center}
\begin{tabular}{cccccc}
\hline
Epoch& Observation &Modified&Number of \\
number& date &Julian date& features\\ \hline
1& 23rd Nov 2008 & 54793 &15\\
2& ~1st Feb 2009 & 54863 &15\\
3& 18th May 2009 & 54969 &10\\
4& 28th Aug 2009 & 55072 &12\\
5& ~15th Sep 2009 $\dag$& 55089 &12&\\
6& ~27th Sep 2009 $\dag$& 55101 &14&\\
7& ~24th Oct 2009 $\dag$& 55128 &13&\\
8& 13th Dec 2009 & 55178 &14\\
9& ~28th Jan 2010 $\dag$& 55224 & -\\
10&10th Feb 2010 & 55237 &12\\
11&~4th Apr 2010 & 55290 &8\\
12&11th Aug 2010 & 55419 &9\\
\hline
\end{tabular}
\begin{tablenotes}
\item $\dag$ These epochs were not used in parallax determination.
\end{tablenotes}
\end{center}
\end{table}


\begin{table*}
\scriptsize
\vspace{-0.2cm}
\begin{center}
\caption{The general properties of H$_{2}$O masers in S255IR-SMA1 detected with VERA. \label{TAB}}
\begin{tabular}{cclcccccccc}
\hline
Maser&$V_{\rm LSR}$&Detected &$\Delta \alpha \cos \delta$&$\Delta \delta$ &$\mu_{\alpha}\cos\delta$&$\mu_{\delta}$&$\pi$\\ 
ID &(km s$^{-1}$)&epochs&(mas)&(mas)&(mas yr$^{-1}$)&(mas yr$^{-1}$)& mas\\
\hline
\textbf{\underline{A}}	&	$	4.798	$	&	12345678*10 11 12	&	26.851	&	-6.846	&	$	1.03	\pm	0.16	$	&	$	-1.01	\pm	0.18	$	&	$0.585 \pm 0.071$	\\
B	&	$	4.166	$	&	12345678*10**	&	8.894	&	-15.367	&	$	-1.31	\pm	0.18	$	&	$	-2.83	\pm	0.16	$	&		\\
C	&	$	4.374	$	&	12345678*10 11 12	&	5.322	&	-12.98	&	$	-1.93	\pm	0.21	$	&	$	-2.10	\pm	0.20	$	&		\\
D	&	$	5.428	$	&	12345678*10 11 12	&	4.018	&	-11.924	&	$	-1.27	\pm	0.20	$	&	$	-0.09	\pm	0.88	$	&		\\
E	&	$	3.745	$	&	*****678*10**	&	0.604	&	-11.377	&	$	-2.30	\pm	0.17	$	&	$	-1.45	\pm	0.15	$	&		\\
F	&	$	5.851	$	&	***456*8*10**	&	-0.87	&	-6.76	&	$	-1.79	\pm	0.20	$	&	$	-1.70	\pm	0.20	$	&		\\
\textbf{\underline{G}}	&	$	0.161	$	&	12345678*10 11 12	&	0.046	&	-4.794	&	$	-2.95	\pm	0.16	$	&	$	-0.13	\pm	0.15	$	&$0.505 \pm 0.100$\\
\textbf{\underline{H}}	&	$	3.953	$	&	12345678*10 11 *	&	-1.019	&	-3.62	&	$	-1.97	\pm	0.18	$	&	$	-1.55	\pm	0.19	$	&$0.502 \pm 0.125$\\
I	&	$	1.846	$	&	1234********	&	-4.352	&	2.585	&	$	-2.73	\pm	1.38	$	&	$	-0.90	\pm	0.71	$	&		\\
\textbf{\underline{J}}	&	$	11.322	$	&	*2345678*10 11 12	&	73.987	&	131.344	&	$	-1.20	\pm	0.17	$	&	$	0.49	\pm	0.17	$	&$0.603 \pm 0.077$\\
K	&	$	10.275	$	&	***45678*10 * 12	&	85.739	&	152.066	&	$	-1.42	\pm	0.16	$	&	$	0.05	\pm	0.15	$	&		\\
\textbf{\underline{L}}	&	$	12.169	$	&	12345678*10 11 12	&	88.909	&	152.196	&	$	-0.95	\pm	0.17	$	&	$	0.27	\pm	0.16	$	&$0.561 \pm 0.060$\\
M	&	$	11.539	$	&	*****678*10 11 12	&	150.041	&	253.838	&	$	-2.95	\pm	0.16	$	&	$	1.22	\pm	0.29	$	&		\\
N	&	$	6.476	$	&	**345*78*** 12	&	275.557	&	245.008	&	$	1.83	\pm	0.19	$	&	$	1.47	\pm	0.25	$	&		\\
O	&	$	5.849	$	&	12**********	&	249.885	&	235.729	&	$	-			$	&	$	-			$	&		\\
P	&	$	3.11	$	&	12**********	&	43.809	&	8.626	&	$	-			$	&	$	-			$	&		\\
Q	&	$	3.321	$	&	12**5678****	&	40.716	&	5.044	&	$	0.51	\pm	0.16	$	&	$	-2.83	\pm	0.45	$	&		\\
R	&	$	2.689	$	&	1***********	&	83.969	&	29.766	&	$	-			$	&	$	-			$	&		\\
S	&	$	3.11	$	&	1***********	&	76.449	&	22.751	&	$	-			$	&	$	-			$	&		\\
T	&	$	0.998	$	&	*2**********	&	1.52	&	39.99	&	$	-			$	&	$	-			$	&		\\
U	&	$	5.217	$	&	1***********	&	22.615	&	-11.963	&	$	-			$	&	$	-			$	&		\\
V	&	$	5.212	$	&	*2**********	&	18.842	&	-15.06	&	$	-			$	&	$	-			$	&		\\
W	&	$	4.585	$	&	1***********	&	16.933	&	-15.294	&	$	-			$	&	$	-			$	&		\\
X	&	$	3.324	$	&	*****6******	&	-2.696	&	-5.001	&	$	-			$	&	$	-			$	&		\\
Y	&	$	13.008	$	&	*2**********	&	173.378	&	264.51	&	$	-			$	&	$	-			$	&		\\
\hline
Systemic &&&&& $-0.22\pm0.19$ & $0.03\pm0.25$ & $0.563 \pm 0.036$\\
\hline
\end{tabular}
\begin{tablenotes}
\item{\footnotesize{Column (2)}: Line of sight velocities are quoted as the value measured at the first detection.}
\item{\footnotesize{Column (3)}: Numbers indicate detection in the corresponding epoch, while asterisk represents non-detection.}
\end{tablenotes}
\end{center}
\end{table*}

The a-priori delay tracking models used in correlation were improved upon after basic inspection of fringe map analysis to correct for the maser offset from the delay tracking center. In this step we also refined delay tracking solutions using more accurate antenna positions utilising global positioning system (GPS) measurements which also measured atmospheric water vapour zenith delays at each station \citep{Honma08b} and include fine corrections for the Earth rotation parameters.

In total there were 12 observation epochs with intended spacing of around 3 months. In three epochs, toward the middle of the observing calendar, we experienced antenna trouble at at least one station - rendering phase referencing between the quasar and maser impossible. Data from these epochs were not used for parallax measurement, but their self-calibrated maps and fluxes contributed to proper motion and spectrum analyses. 
Epoch 9 sustained a fatal error and subsequently could not be imaged, only its spectrum was useful. The observation calendar is summarised in Table~\ref{obs}.

All data were reduced using the Astronomical Image Processing System (AIPS) developed by the National Radio Astronomy Observatory (NRAO).
Data reduction utilised the \emph{inverse phase-referencing} method for VERA data, which was introduced in \citet{Imai12}. A detailed guide is given in \citet{Burns15a}. Flux calibration was performed using system temperatures and gain information recorded at each station. Next we applied the aforementioned delay tracking solutions which are source specific. 
A characterised noise signal was injected into the two VERA beams to allow correction of the delay introduced by differences of hardware in the dual beam system \citep{dual}, and group delay was calibrated using the intermittent scans of BLLAC, DA55 or 3C84. What remains at this point are time-varying phase residuals attributed to atmospheric fluctuations which, at 22 GHz, are  dominated by the dynamic troposphere \citep{Asaki07}. Atmospheric calibration was carried out by fringe fitting phase residuals using the emission from a strong reference water maser at solution intervals of 1 or 2 minutes. These solutions were then applied to the visibility data of J0613+1708, phase-referencing it. Imaging of J0613+1708 then gives the relative astrometric position of water masers in S255IR-SMA1 with respect to J0613+1708. 
The phase solutions obtained from the reference maser channel were also applied to all other maser channels, achieving rms noise values of typically 100 - 300 mJy beam$^{-1}$.

Maser maps were produced by applying the CLEAN procedure, based on \citet{Hog74}, to emission peaks registered at a signal-to-noise cutoff of 7. Following common nomenclature, a maser `spot' refers to an individual maser brightness peak, imaged in one spectral channel, and a maser `feature' refers to a group of spots which are considered to emanate from the same physical maser cloud, thus a maser feature typically comprises of several maser spots. Maser spots are categorised into features when they are part of the same spectral feature and found within 1 mas of another spot in that feature. We define the nominal astrometric position of a maser feature by determining the flux weighted average of the brightest three spots in the feature. 



\begin{figure}
\begin{center}
\includegraphics[scale=1.15]{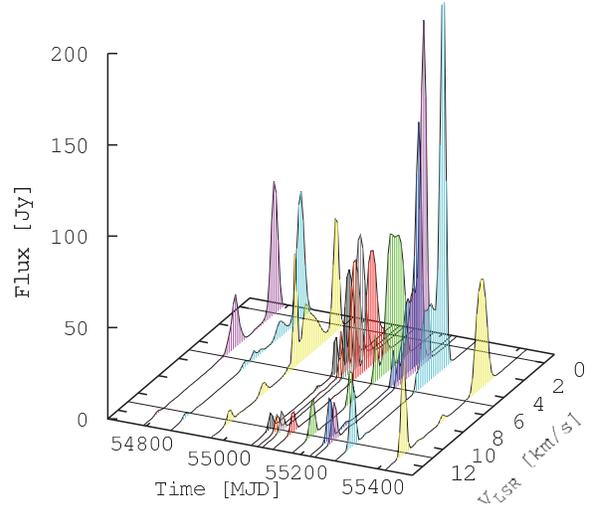}
\caption{Scalar averaged cross-power spectra of left-hand circularly polarised maser emission in S255IR-SMA1 as a function of time. Colours are arbitrary. 
\label{spectra}}
\end{center}
\end{figure}

\begin{figure}
\begin{center}
\includegraphics[scale=1.08]{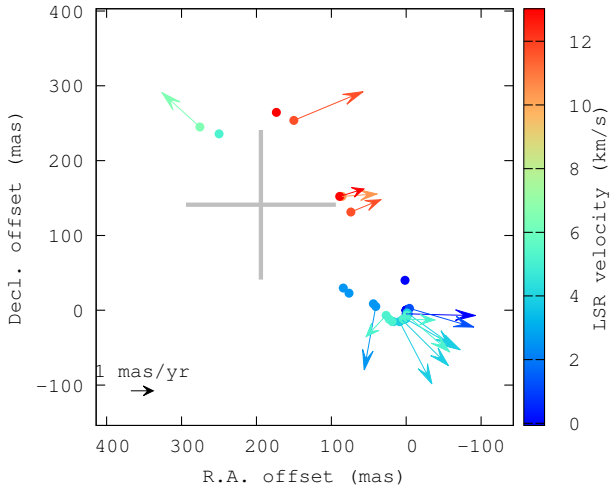}
\caption{Distributions, internal proper motions and line-of-sight velocities of H$_2$O maser features in S255IR-SMA1. The peak position of the centimeter source from \citet{Reng96} is indicated with a cross whose size indicates to the positional error in the measurement.
\label{jet}}
\end{center}
\end{figure}

\section{Results}

\subsection{General properties of masers}

In total, 25 individual maser features were detected with VERA, each comprising of multiple maser spots. The LSR velocities, detection frequency, positions and proper motions of all maser features are summarised in Table~\ref{TAB} (positions are quoted as offsets from the phase tracking center in the maser beam, \emph{see} Section 2). The time-evolution of the maser spectrum is shown in Figure~\ref{spectra}, and shows some velocity features to be stable while others are more variable.

Feature J exhibited a steady increase in flux from 4 to 82 Jy which can be seen clearly in Figure~\ref{spectra} at $v_{\rm LSR} = 11.3$ km s$^{-1}$, while conversely, Feature G at $v_{\rm LSR} = 0.2$ km s$^{-1}$ exhibited a gradual weakening. These changes in flux were not accompanied by any change in the structure of the masers in our images. As such the slow flux changes may be attributed to some physical change in the environment near S255IR-SMA1; a change in the intensity of seed photons or a change in the density or path length of the maser cloud. Presently we cannot speculate beyond this.

We observed a chance alignment where the spectral peaks of two bright maser features (Features C and G) drifted into a common velocity channel. This observational- rather than  physical effect fully accounts for the flux increase seen at about $MJD=55230$ days in Figure~\ref{spectra}.

\begin{figure}
\begin{center}
\includegraphics[scale=0.77]{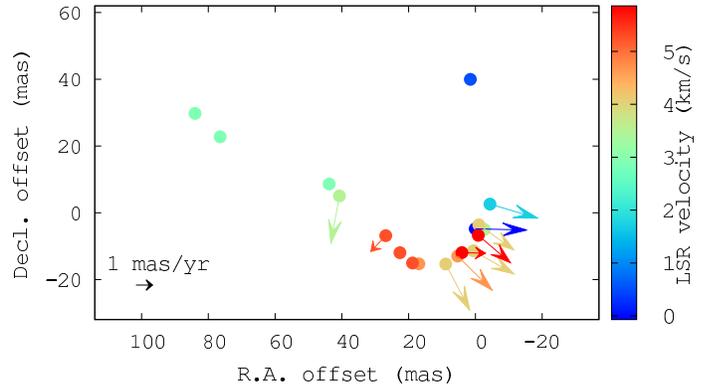}
\caption{Zoom of the SW maser jet in S255IR-SMA1. The distributions and motions of masers delineate a U-shaped bowshock.
\label{micro}}
\end{center}
\end{figure}

The distribution of water maser features in S255IR-SMA1 is shown in Figures~\ref{jet} and~\ref{micro}. Masers form four main groups; those located to the NE of the system which are slightly redshifted with respect to the star ($V_{\rm LSR} =+5.25$ km s$^{-1}$, based on the central velocity of the rotating core observed by \citealt{Zin15}), those in the SW which are slightly blueshifted with respect to the star, and two groups of redshifted water masers ($\sim$ 12 km s$^{-1}$) North and West of the source of the centimeter emission marked as a plus sign in Figure~\ref{jet}. We refer to these four maser groups as the `NE masers', the `SW masers', the `red-N masers' and the `red-W masers', respectively.





\subsection{Annual parallax}
\label{pie}

The annual parallax of a maser source, obtained via astrometry, can be used to measure its trigonometric distance.
Astrometric motions of maser features trace curlicue paths across the sky-plane. This motion can be separated into a linear component which arises from the proper motion of the maser with respect to the Sun, and a sinusoidal component caused by the annual parallax. Astrometric maser motions are deconstructed into these separate components by simultaneous fitting of data with a linear and a sinusoidal function, assuming a common distance. We only performed this fitting for maser features which were observable in at least 7 epochs, spanning at least one year, and that were not spatially resolved. Maser features suitable for fitting are indicated by underlined, boldface feature ID's in Table~\ref{TAB}.
We avoided using features C and D, eventhough they were present in 11 epochs, because spatial interaction of these two features, with each other, distorted the maser images - compromising astrometric accuracy.

\begin{figure}
\begin{center}
\includegraphics[scale=1.9]{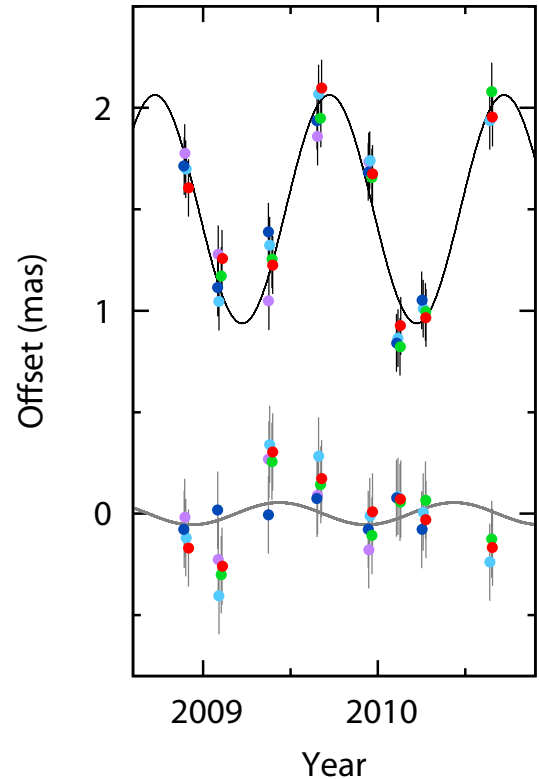}
\caption{Annual parallax motions of 5 maser features in S255IR-SMA1 which are coloured arbitrarily. The upper, black line shows the modelled parallax in the Right ascension direction while the grey line below shows the same for the declination direction. Small horizontal offsets were introduced for readability.
\label{squiggle}}
\end{center}
\end{figure}

Simultaneously fitting 5 maser features we measured an annual parallax of $\pi = 0.563 \pm 0.036$ mas (Figure~\ref{squiggle}), corresponding to a source distance of D $= 1.78^{+0.12}_{-0.11}$ kpc. The parallaxes of individual maser features are shown in column 8 of Table~\ref{TAB} - their self-consistency support the validity of the simultaneous fit.

\citet{Rygl10} measured the annual parallax of S255 to be $\pi = 0.628 \pm 0.027$ mas, using 6.7 GHz methanol masers and the European VLBI Network (EVN); ours and their estimates differ by more than one sigma. Though their estimate reports a higher precision it is likely that our estimate is the more accurate for several reasons: water masers are brighter and more compact, thus providing better astrometric accuracy, than 6.7 GHz methanol masers; our parallax fitting uses 8 epochs, compared to the 4 epochs in their estimate; and our estimate utilises 5 maser features, compared to the 1 in their estimate - thus we can expect a better reduction of random errors.



\subsection{Systemic and internal proper motion}
\label{intMots}

We measured the proper motions of maser features that were detected in at least 3 epochs by measuring motions relative to Feature G (the reference maser) in the self-calibrated maps. The absolute proper motion of Feature G, obtained from the parallax fitting stage outlined in Section~\ref{pie}, was then added to all other masers, thus converting proper motions relative to Feature G into absolute proper motions relative to the reference source, J0613+1708.
Errors were calculated in quadrature to include the uncertainty in the motion of Feature G.


The observed proper motions of maser features are a combination of two components; the `source systemic motion' which is a group motion common to all masers arising from the relative motion of the target and the observer, and `internal motions' which are motions of individual maser features with respect to the driving source. 
In S255IR-SMA1 we were able to measure proper motions in both the NE and the SW masers.
First we averaged the motions of maser spots associated with each outflow lobe to get the individual lobe proper motions. Then we determined the systemic motion from the residual of summing the proper motions of the lobes, assuming symmetry and with equal weighting.
We did not include the red masers in this calculation as they do no appear to be associated with the symmetric outflow system.
We derive a (heliocentric) systemic motion of ($\mu_{\alpha}\cos\delta$, $\mu_{\delta}$) = ($ -0.13\pm0.20$, $-0.06\pm0.27$) mas yr$^{-1}$, where errors are the quadrature sum of the average error for each lobe, divided by $\sqrt{N}$ where $N$ is the number of features in the lobe. 

Our result is consistent with the estimate of \citet{Rygl10} of ($\mu_{\alpha}\cos\delta$, $\mu_{\delta}$) = ($ -0.14\pm0.54$, $-0.84\pm1.76$) mas yr$^{-1}$, using 6.7 GHz methanol masers which are considered to be good tracers of systemic motions since they are excited radiatively at close proximity to the star. Our higher precision likely comes from the symmetry of the detected water maser kinematics and our use of many maser features. Comparing S255IR-SMA1 with nearby Perseus Arm maser sources: S252 ($\mu_{\alpha}\cos\delta$, $\mu_{\delta}$) = ($+0.02\pm0.30$, $-2.02\pm0.30$) mas yr$^{-1}$ \citep{Reid09a,Oh10} and G192.16-3.84 ($\mu_{\alpha}\cos\delta$, $\mu_{\delta}$) = ($+0.69\pm0.05$, $-1.57\pm0.15$) mas yr$^{-1}$ \citep{Shio11}, the difference is small enough to be attributed to peculiar motion with regards to Galactic rotation.

Subtracting our systemic proper motion from the measured proper motions gives the internal motions of water masers with respect to the source. These are shown in Figures~\ref{jet} and~\ref{micro}, and reveal a highly collimated bipolar jet orientated in the NE-SW direction ($\mathrm{PA} = 49^{\circ}$). 
At our distance, we calculate physical sky-plane maser velocities of $v_{\rm sky} =18.74$ km s$^{-1}$. Comparing masers in the NE and SW lobes we note that $\Delta v_{\rm sky} > \Delta v_{\rm LOS}$ meaning that the 3D motion vectors are dominated by proper motions i.e. the jet moves primarily in the sky-plane (we calculate the jet's inclination as $i=86.25^{\circ}$ to the observer). The 3D velocity of each jet lobe is $v_{\rm 3D} =18.78$ km s$^{-1}$ with respect to the protostar, with a position angle of $\mathrm{PA} = 49^{\circ}$ which matches reasonably well with those of the larger molecular outflows ($\mathrm{PA} = 67^{\circ}$, \citealt{Zin15}); perpendicular to the rotating core/disk.

Masers appear to trace a shock front propagating away from the centimeter source in the SW direction. Maser-traced bowshock structures of similar appearance have been reported in the literature including the low mass YSO S106-FIR \citep{Furuya00} and also in high mass YSOs AFGL2591 \citep{Sanna12,Trinidad13} and G31.41+0.31 \citep{Moscadelli13}. Such formations are thought to arise in the shocked gas at the interface between a protostellar jet and its surroundings. We return to this topic in Section~\ref{JETMODEL}.


\begin{figure}
\begin{center}
\includegraphics[scale=0.44]{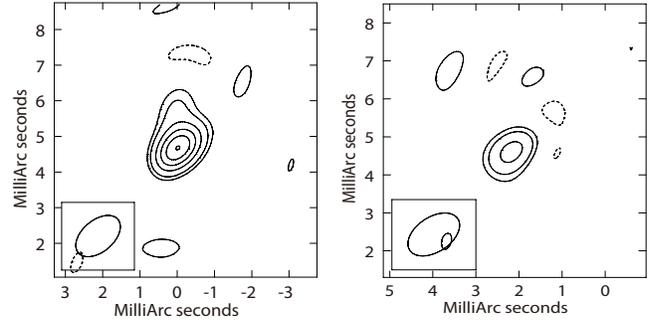}
\caption{J0613+1708 images from (\emph{left}) epoch 1, in which good observing conditions were seen at all VLBI stations, and in (\emph{right}) epoch 4, in which conditions were not so good. The contours correspond to -3,3,5,9,13,17,21 times the image \emph{rms} noise which are 1.8 mJy and 2.8 mJy for epochs 1 and 4 respectively.
\label{QuasarJet}}
\end{center}
\end{figure}

\begin{figure*}
\begin{center}
\includegraphics[scale=2.2]{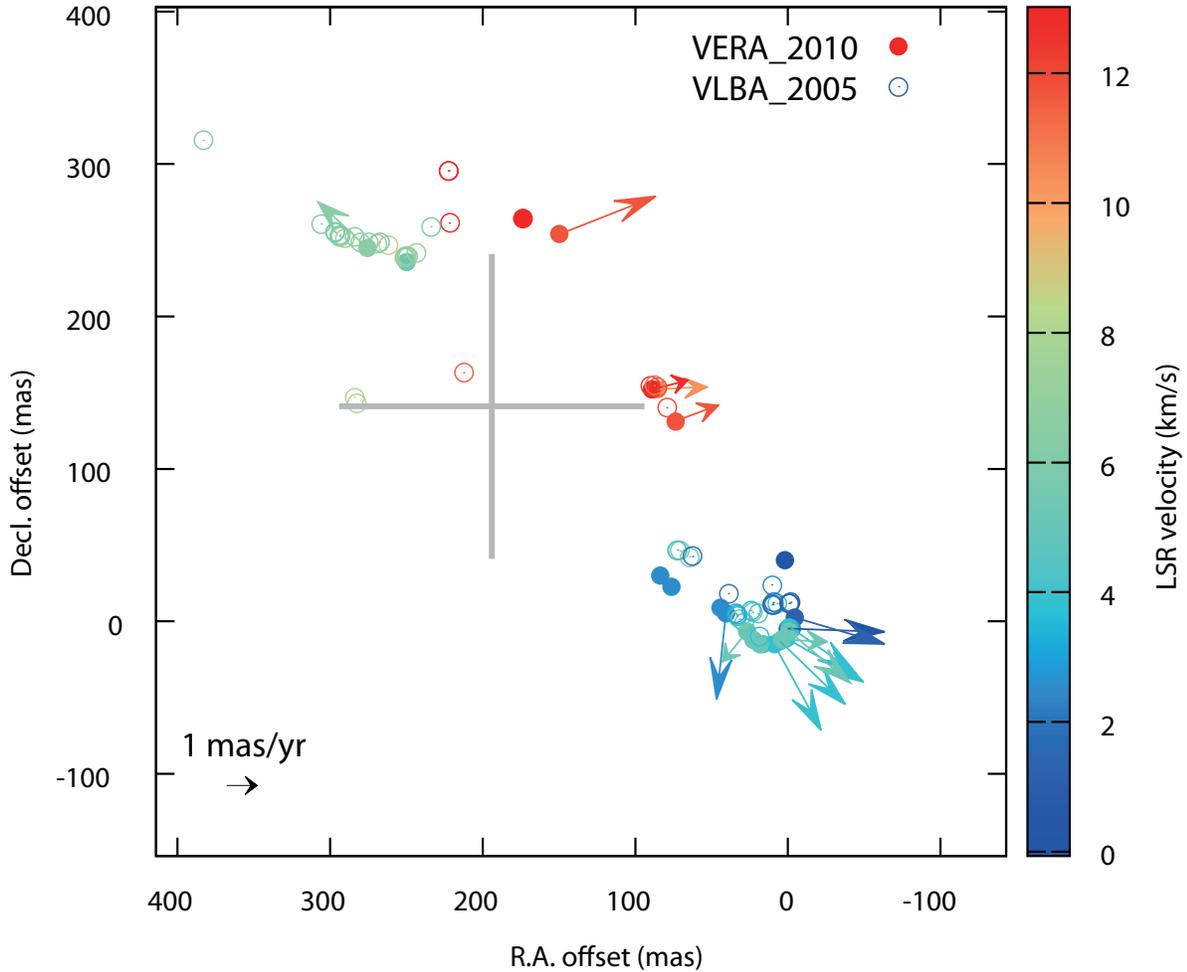}
\caption{Distributions of water masers observed in 2005 using the VLBA \citep{Goddi07}, and in 2010 using VERA (this work). Masers from the aforementioned works are shown as open and filled circles, respectively. The VLBA masers were shifted into the frame of the MYSO by correcting for the systemic motion over the 5 years elapsed between observations. The peak position of the centimeter source from \citet{Reng96} is indicated with a cross whose size indicates to the positional error in the measurement.
\label{GODDIROSS}}
\end{center}
\end{figure*}

\subsection{Quasar jet in J0613+1708}


The reference source J0613+1708 has a quasar jet in the North direction which seen at in K-band continuum (Figure~\ref{QuasarJet}) extending to about 2 mas from the position of the core. This jet structure is seen more clearly in X-band images from the Second VLBA Calibrator Survey: VCS2 \citep{VLBA2}. In the X-band image the jet extends about 10 mas from the core. 

Non-point-like structures in the positional reference source, such as those arising from a quasar jet, can influence the 2D Gaussian fitting used in determining its position from the interferometric images - in turn degrading maser astrometry. We checked to see if such an issue affected our data.


The extent of the jet in our images depended on the rms noise suppression achieved after inverse phase referencing - which differs for each observing epoch depending on observing conditions; in some epochs the jet is clearly resolved while in others the jet is buried in noise (Figure~\ref{QuasarJet}). 
An influence on the astrometric accuracy, if present, would manifest as a larger parallax fitting residuals in the Dec. direction; the direction of the jet. Yet, since the fitting residuals in R.A. and Dec. were similar (about 0.1 mas in both cases) it seems that the jet structure did not significantly degrade the astrometric accuracy of our observations.

\newpage

\section{Discussion}

\subsection{Combined VERA and VLBA view of S255IR-SMA1}

A more complete view of the maser activity in S255IR-SMA1 can be obtained by comparison with the previous observational results of \citet{Goddi07} who observed the same maser transition using the VLBA. Their observations were carried out in 2005, and ours in 2010. We combine our maser data with the positions and line of sight velocities of the 56 masers in their Table 1, corrected for the offset caused by the source motion during the 5 years between observations. The comparison is shown in Figure~\ref{GODDIROSS}.

Although individual maser features are unlikely to have survived the time between VLBA and VERA observations it is clear that the physical structures traced by the different maser groups (NE, SW and red-W masers) have persisted. Compared to the VERA only map (Figure~\ref{jet}) the combined maser distribution better samples the physical structure and reinforces the NE and SW lobes as dominant sources of maser emission. A lengthening of the jet is evident in the SW lobe, the case is not so clear in the NE lobe as no masers sample the shock front.

An expanded view of the SW jet head is given in Figure~\ref{GODDIROSSMOD} which shows jet lengthening and tentative indication of expansion in a jet-widening sense. This is difficult to quantify since the VLBA data contains two incomplete possible shock fronts, making cross identification with the VERA results difficult. We return to this topic, and that of jet widening in Section~\ref{JETMODEL}, via comparison to jet models.

From the combined VERA and VLBA view we infer a jet shell width of about 50 mas. The angular separation between the SW and NE lobes is about 400 mas, leading to a collimation degree of about 8. Taking in to account the system inclination of $i=86.25^{\circ}$ and distance of 1.78 kpc, we estimate a physical jet length of $\sim 335$ au. 
Using our 3D maser velocity we calculate a dynamic age of $t_{\rm jet} \leq 130$ years. Since the jet bowshock moves faster than the gas it entrains, the maser motions represent a lower limit to the jet velocity and thus the dynamic age is an upper limit. Given that protostellar ejection are thought to follow accretion events the occurrence of a young bipolar jet implies that S255IR-SMA1 is still actively accreting mass.




\subsection{Comparison with jet- and wind-driven models}
\label{JETMODEL}

\begin{figure*}
\begin{center}
\includegraphics[scale=0.93]{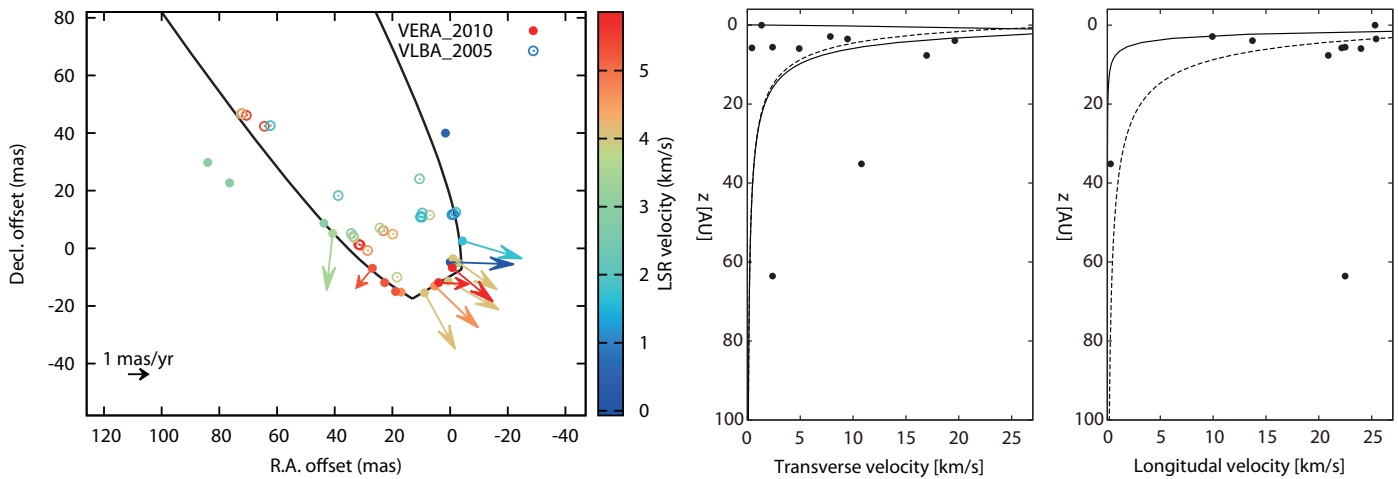}
\caption{\emph{Left} shows masers associated with the SW bow shock from the 2005 VLBA observations of \citet{Goddi07} (empty circles), and the 2010 VERA observations from this work (filled circles) where arrows indicate proper motions (from this worm only). Overlain is the locus of jet shell model from \citet{Ostriker01} calculated for a primary jet radius of $R_j = 6$ AU. \emph{Middle} and \emph{right} respectively show the transverse and longitudinal proper motion velocities expected for gas at distances along the jet, z, where the bowshock head is at z $=0$ AU. Functions represent material newly swept up by the propagating jet (solid line) and the average motion of the jet's outer shell (dashed line) from the model. Overlain to these are the measured transverse and longitudinal proper motions measured by VERA.
\label{GODDIROSSMOD}}
\end{center}
\end{figure*}


The elongated NE-SW morphology and bipolar outward motions of water masers in
S255IR-SMA1 suggest that they are associated with an ejection event from a MYSO.
Bipolar outflow- and jet-tracing maser observations are frequently reported in the literature (\emph{ex.} \citealt{Imai07,NY08,Moscadelli11,Torrelles14} and G236.81+1.98 in \citealt{Choi14}), such results reveal the physical shape, scale and kinematics (if measured) of the shocked gas shell at the interface between the outflowing and ambient gasses - allowing a comparison with physical models.

We compared maser distributions and proper motions in the SW maser lobe to the models of \citet{Ostriker01} and \citet{Lee01} who, in a pair of papers, describe and compare the jet- and wind-driven outflows of protostars. Figure~\ref{GODDIROSSMOD} (\emph{left}) shows that the distribution of water masers observed by VERA is remarkably well matched by the locus of the jet driven bow shock described by Equation 22 of \citet{Ostriker01}, using a jet radius of 6 AU and position angle of $49^{\circ}$ based on the maser data.

Figure~\ref{GODDIROSSMOD} (\emph{middle}) shows the predicted proper motions of masers in the direction transverse to the jet. Functions are those predicted by the models of \citet{Ostriker01} (their Equations 19 and 21), in which we adopt their values for: the isothermal sound speed $c_s = 8$ km s$^{-1}$, and the ratio of the velocity of material being ejected from the jet surface to the velocity of the propagating bowshock $\beta c_s / v_s = 0.5$. Our singular input parameter was jet radius which we set to $R_j=6$ AU based on the match to our maser data. 

In the case of a bow shock, gas is forced edgeways at the head of the forward-propagating jet resulting in large transverse velocities in addition to small transverse velocities associated with the forward-propagating gas. Generally, Figure~\ref{GODDIROSSMOD} (\emph{middle}) can be understood as a plot of these transverse velocities at various distances from the head of the bow shock located at z $=0$ AU. Further from the influence of the bow shock, gas only moves transverse to the jet direction by thermal expansion - we thus see a sharp fall in the transverse velocity away from the jet head. 

Our maser data show a large dispersion of transverse velocities near the jet head (Figure~\ref{GODDIROSSMOD}, \emph{middle}) as is expected from the jet-driven outflow model. In contrast, a wind-driven outflow model predicts zero transverse velocity at the jet head while the largest transverse velocities are expected about half-way along the body of the outflow (Figure 9 of \citealt{Lee01}). 


Two masers show unexpectedly large transverse velocities considering their distances from the jet head (at $z = 35$ and 64 AU). Consequently an additional source of expansion, such as contribution from a disk wind, must be invoked to explain how these masers reached transverse velocities of up to 10 km s$^{-1}$. This can be reconciled if the outflow is powered by a combination of jet- and wind-driven mechanisms as opposed to a purely jet driven scenario. A similar situation was reported by \citet{Sanna12}, observing water masers in AFGL 2591.

A transverse expansion velocity of 10 km s$^{-1}$ over 5 years would produce an offset of 6 mas which is comparable to the separation between the jet-shell model and the positions of the outermost (from the jet axis) masers observed by \citet{Goddi07} (see Figure~\ref{GODDIROSSMOD}). Thus the outermost VLBA masers likely trace the same (expanding) surface as the VERA masers. The innermost masers (closer to the jet axis) represent a secondary surface. The separation of this surface from the shock front observed by VERA is 20 mas. An internal motion of 32 km s$^{-1}$ over 5 years would be required to explain these as a common shock front - far larger than the fastest measured internal motion in the jet direction ($v_{\rm z max} = 25$ km s$^{-1}$). Thus we deem the inner VLBA maser surface to be independent of the VERA maser surface.

The two surfaces may represent internal and external surfaces in a single bowshock; the inner surface may represent the jet shock while the outer surface may represent the bow shock created in front of the jet as it sweeps up ambient gas. In such a case we calculate the `thickness' of the bow shock as the current separation of the surfaces (36 AU), minus the internal motion traversed in 5 years (26 AU), giving 10 AU, which is comparable to the jet width.
Alternatively the second surface may represent a second shock front from a repeating jet. In this case, the time between ejections would be about 3 yrs, assuming constant jet velocities of $v_s = 66$ km s$^{-1}$, based on the model of \citet{Ostriker01}. The required interval between ejections is shorter than the tens to a few hundred years for which accretion bursts themselves are thought to last, making a 3 yr repeating jet scenario unfeasible. This leads us to favour the former interpretation - that masers trace two surfaces in a single jet bowshock.

Figure~\ref{GODDIROSSMOD} (\emph{right}) represents the predicted proper motions in the jet direction. Again, the water masers observed with VERA are well described by the jet-driven outflow model of \citet{Ostriker01} (Equations 18 and 20). And again data contrasts that expected for a wind-driven outflow (a linear relation between velocity and z, \emph{see} Figure 9 of \citealt{Lee01}). Hydrodynamic simulations of \citet{Lee01} show that the highest velocity gas associates with the jet shell (dashed line in our figure) while the lower velocity gas is attributed to ambient material that has been entrained. In agreement, it can be seen that masers with the highest measured longitudinal velocity lie closely to the jet outer shell. These masers are therefore shocked on the boundary between the jet and the ambient material - corroborating the spatial association of maser positions to the jet shell model seen in Figure~\ref{GODDIROSSMOD} (\emph{left}).
One maser has a far larger longitudinal velocity than that predicted by the jet model (at z = 64 AU). As before, this behaviour may be explained by invoking a wind component accompanying the jet.


\citet{Goddi07} conclude that the water masers in S255IR-SMA1 likely associate with a disk-wind. Their interpretation was based, similarly to ours, on the distribution and proper motions of water masers. Comparing our proper motions with theirs, consistency is seen in the red masers (the `Aw' cluster in their Figure 1), however they measure fewer proper motions associated with the NE-SW jet components. Our jet-tracing masers show a more systematic expansion motion and give a more complete sampling of the bow shock surface in the SW lobe.
Our analysis of the proper motions of water masers in S255IR-SMA1 support a predominantly jet-driven outflow origin, possibly including contribution from a wind. Our jet radius of $R_j = 6$ AU is similar to the 6 AU used by \citet{Sanna12} for a similar analysis of jet-tracing masers in AFGL 2591. 



\subsection{Episodic ejection}

Another important feature of low mass star formation is episodic accretion in which short intense bursts of accretion onto the central star perforate periods of inactivity lasting a few thousand years - accretion bursts are known from observations \citep{Herbig77,Hartmann96}, and simulations \citep{Zu09,Stamatellos11}.
In the case of massive primordial star formation episodic accretion bursts may play an important role in regulating the ionising radiation emitted from the embedded massive star - providing a mechanism of prolonging accretion into the UCHII region phase of massive star birth \citep{Hosokawa15}. The timescales involved in episodic accretion in very massive stars are simlilar to those of low mass stars, though the proposed purpose of episodic accretion in both cases is different.

With the exception of FU Ori stars, episodic accretion is impractical to observe due to the long timescales involved. However, the episodic nature of accretion in young stellar objects can be inferred from evidence of episodic ejection observable as symmetric pairs of ejection bow shocks at ever increasing distances from the central object (for example, in HH111 \citep{Bo89}).

At least three episodes of ejection are known to have taken place in S255IR-SMA1. The larger, and therefore older outflow was observed in $^{12}$CO$(2-1)$ and has dynamic age of about 7000 years \citep{Wang11}, and a second younger ejection indicated by bow shocks of Fe$\mathrm{II}$ line emission \citep{Wang11} which is also seen in the HCO$^+$ maps of \citet{Zin15} who estimate its dynamic age to be 1000 years. The third being our maser jet bow-shock of dynamic age $t_{dyn}\leq 130$ years. 
We conclude that the 20 M$_{\odot}$ MYSO in S255IR-SMA1 has formed accompanied by three separate episodes of ejection during the last few thousand years. 


\section*{Acknowledgments}

R.B. would like to acknowledge the Ministry of Education, Culture, Sports, Science and Technology (MEXT) Japan for support as part of the Monbukagakusho scholarship.

\def\ref@jnl#1{{\rmfamily #1}}%
\newcommand\aj{\ref@jnl{AJ}}%
\newcommand\araa{\ref@jnl{ARA\&A}}%
\newcommand\apj{\ref@jnl{ApJ}}%
\newcommand\apjl{\ref@jnl{ApJ}}%
\newcommand\apjs{\ref@jnl{ApJS}}%
\newcommand\ao{\ref@jnl{Appl.~Opt.}}%
\newcommand\apss{\ref@jnl{Ap\&SS}}%
\newcommand\aap{\ref@jnl{A\&A}}%
\newcommand\aapr{\ref@jnl{A\&A~Rev.}}%
\newcommand\aaps{\ref@jnl{A\&AS}}%
\newcommand\azh{\ref@jnl{AZh}}%
\newcommand\baas{\ref@jnl{BAAS}}%
\newcommand\jrasc{\ref@jnl{JRASC}}%
\newcommand\memras{\ref@jnl{MmRAS}}%
\newcommand\mnras{\ref@jnl{MNRAS}}%
\newcommand\pra{\ref@jnl{Phys.~Rev.~A}}%
\newcommand\prb{\ref@jnl{Phys.~Rev.~B}}%
\newcommand\prc{\ref@jnl{Phys.~Rev.~C}}%
\newcommand\prd{\ref@jnl{Phys.~Rev.~D}}%
\newcommand\pre{\ref@jnl{Phys.~Rev.~E}}%
\newcommand\prl{\ref@jnl{Phys.~Rev.~Lett.}}%
\newcommand\pasp{\ref@jnl{PASP}}%
\newcommand\pasj{\ref@jnl{PASJ}}%
\newcommand\qjras{\ref@jnl{QJRAS}}%
\newcommand\skytel{\ref@jnl{S\&T}}%
\newcommand\solphys{\ref@jnl{Sol.~Phys.}}%
\newcommand\sovast{\ref@jnl{Soviet~Ast.}}%
\newcommand\ssr{\ref@jnl{Space~Sci.~Rev.}}%
\newcommand\zap{\ref@jnl{ZAp}}%
\newcommand\nat{\ref@jnl{Nature}}%
\newcommand\iaucirc{\ref@jnl{IAU~Circ.}}%
\newcommand\aplett{\ref@jnl{Astrophys.~Lett.}}%
\newcommand\apspr{\ref@jnl{Astrophys.~Space~Phys.~Res.}}%
\newcommand\bain{\ref@jnl{Bull.~Astron.~Inst.~Netherlands}}%
\newcommand\fcp{\ref@jnl{Fund.~Cosmic~Phys.}}%
\newcommand\gca{\ref@jnl{Geochim.~Cosmochim.~Acta}}%
\newcommand\grl{\ref@jnl{Geophys.~Res.~Lett.}}%
\newcommand\jcp{\ref@jnl{J.~Chem.~Phys.}}%
\newcommand\jgr{\ref@jnl{J.~Geophys.~Res.}}%
\newcommand\jqsrt{\ref@jnl{J.~Quant.~Spec.~Radiat.~Transf.}}%
\newcommand\memsai{\ref@jnl{Mem.~Soc.~Astron.~Italiana}}%
\newcommand\nphysa{\ref@jnl{Nucl.~Phys.~A}}%
\newcommand\physrep{\ref@jnl{Phys.~Rep.}}%
\newcommand\physscr{\ref@jnl{Phys.~Scr}}%
\newcommand\planss{\ref@jnl{Planet.~Space~Sci.}}%
\newcommand\procspie{\ref@jnl{Proc.~SPIE}}%

\bibliographystyle{mn2e}
\bibliography{Clean_R2_IC2162.bib}

\label{lastpage}

\end{document}